\documentclass[twocolumn,aps,prl,preprint,11pt]{revtex4}
\usepackage{graphicx}
\usepackage{epsfig}
 
\newcommand{\be}{\begin{equation}}
\newcommand{\ee}{\end{equation}}

\begin{document}


\title{Mat. Res. Soc. Symp. Proc. Vol. \bf {718} (Perovskite Materials) 25-33 
(2002)\\ \vspace{.2 in} Structural Aspects of Magnetic Coupling in 
     CaCu$_3$Mn$_4$O$_{12}$ and CaCu$_3$Ti$_4$O$_{12}$ }
\author{M. D. Johannes,$^1$ W. E. Pickett$^1$, and R. Weht$^{2}$}

\affiliation{$^1$Department of Physics, University of California, 
          Davis CA 95616}
\affiliation{$^2$Departamento de F\'{\i}sica, CNEA, Avda. General Paz y
   Constituyentes,\\
       1650 - San Mart\'{\i}n, Argentina}
%
\begin{abstract}
Two perovskite-derived materials, CaCu$_3$Mn$_4$O$_{12}$  and  
CaCu$_3$Ti$_4$O$_{12}$, have drawn much recent interest due to their 
magnetoresistive,
dielectric, and magnetoelectronic characteristics.  Here we present
initial theoretical insights into each of these points, based on first
principles, density functional based calculations.  Our results predict CCMO
to have a spin-asymmetric energy gap, which leads to distinct temperature-
and magnetic field-dependent changes in properties, and helps to account
for its observed negative magnetoresistivity.  We have studied CCTO
primarily to gain insight into the exchange coupling in both these
compounds, where the conventional superexchange coupling vanishes by
symmetry for both nearest and next nearest Cu-Cu neighbors, a consequence
of the structure.  In CCTO,
it is necessary to go to 5th Cu-Cu neighbors to obtain a (superexchange)
coupling that can provide the coupling necessary to give three dimensional
order.  Non-superexchange mechanisms may be necessary to describe the
magnetic coupling in this structural class.
\end{abstract}
\maketitle

\section{Introduction}

The mineral CaTiO$_3$, discovered in 1839 in the Ural Mountains and named
after Count von Perovski, is the prototype of what has since been found
to be a large class of ABO$_3$ perovskite materials.
The ideal structure has a simple
cubic Bravais lattice with A atoms on the corners, B atoms at the center, and 
oxygens on the face.  Alternatively, the structure is made up of BO$_6$ 
corner-shared octahedra, with the A atom placed in the interstitial regions
between eight octahedra.  This latter picture of corner-shared octahedra
has strong theoretical backing and great structural utility, because many of
the structural variations observed in perovskites can be characterized simply
as collective rotations (or at high temperature, incoherent ones) of the 
octahedra.  

Here we will address a class with a quadrupled perovskite structure,
in which four ABO$_3$ units comprise the primitive cell.\cite{nomenclature}
This structural class \cite{chenavas} can be obtained from the 
ideal structure (1) by 
replacing three out of every four A site ions (nearly always closed shell
ions) with an A$^{\prime}$ ion, which quadruples the cell to 
AA$^{\prime}_3$B$_4$O$_{12}$, and then (2)
performing a correlated rotation of the four octahedra, each around one of
the $<111>$ axes, until the A$^{\prime}$ ion is fourfold 
coordinated with O ions 
in a nearly square arrangement.
This tilting leaves one 
triangular face of each octahedra perpendicular to a (111) axis of the 
cubic lattice (Fig. 1).  The need for a Jahn-Teller ion as the 
A$^{\prime}$ ion was
recognized early on; especially for A$^{\prime}$ = Cu$^{2+}$, this distortion
leaves it with the square planar configuration it commonly demands.
The Mn$^{2+}$ ion is also found in the A$^{\prime}$  position in this structure.

In this paper we will address the electronic and magnetic structure of two
of these compounds, the hybrid cupromanganite CaCu$_3$Mn$_4$O$_{12}$ (CCMO) and
the hybrid cuprotitanate CaCu$_3$Ti$_4$O$_{12}$ (CCTO).  They are isostructural,
with the CuO$_4$ plaquettes very nearly square -- the bonds connecting Cu 
and O are all the same length, though the angles between bonds deviate 
slightly from 90 degrees.  Similarly, the distances between the B-site ion 
(Mn or Ti) and each of its six neighbors are identical, but the angles differ 
slightly from 90$^{\circ}$.   
Each oxygen now belongs to a single planar CuO$_4$ plaquette and to two tilted 
(MnO$_6$ or TiO$_6$) octahedra.

The additional chemical as well as structural complexity of this quadrupled
structure results in behavior that is different than in ABO$_3$ perovskites,
and some of this behavior may be unique due to the specific stereochemical
relationships in this class of materials.  The Cu$^{2+}$ ion in this 
structure has been discussed in a Jahn-Teller context by Lacroix\cite{lacroix},
however, it is so far removed in this structure from quasi-cubic symmetry that
this viewpoint is limited in utility.  The necessity of a Jahn-Teller ion in
this site strongly promotes stoichiometry, certainly in the Cu site but
apparently overall -- no significant deviation from stoichiometry could be
detected in  CaCu$_3$Ti$_4$O$_{12}$ and CdCu$_3$Ti$_4$O$_{12}$, where 
thorough structural studies have been possible.\cite{exp,cc,mas}

\begin{figure}[tbp]
\includegraphics[width=2.3in]{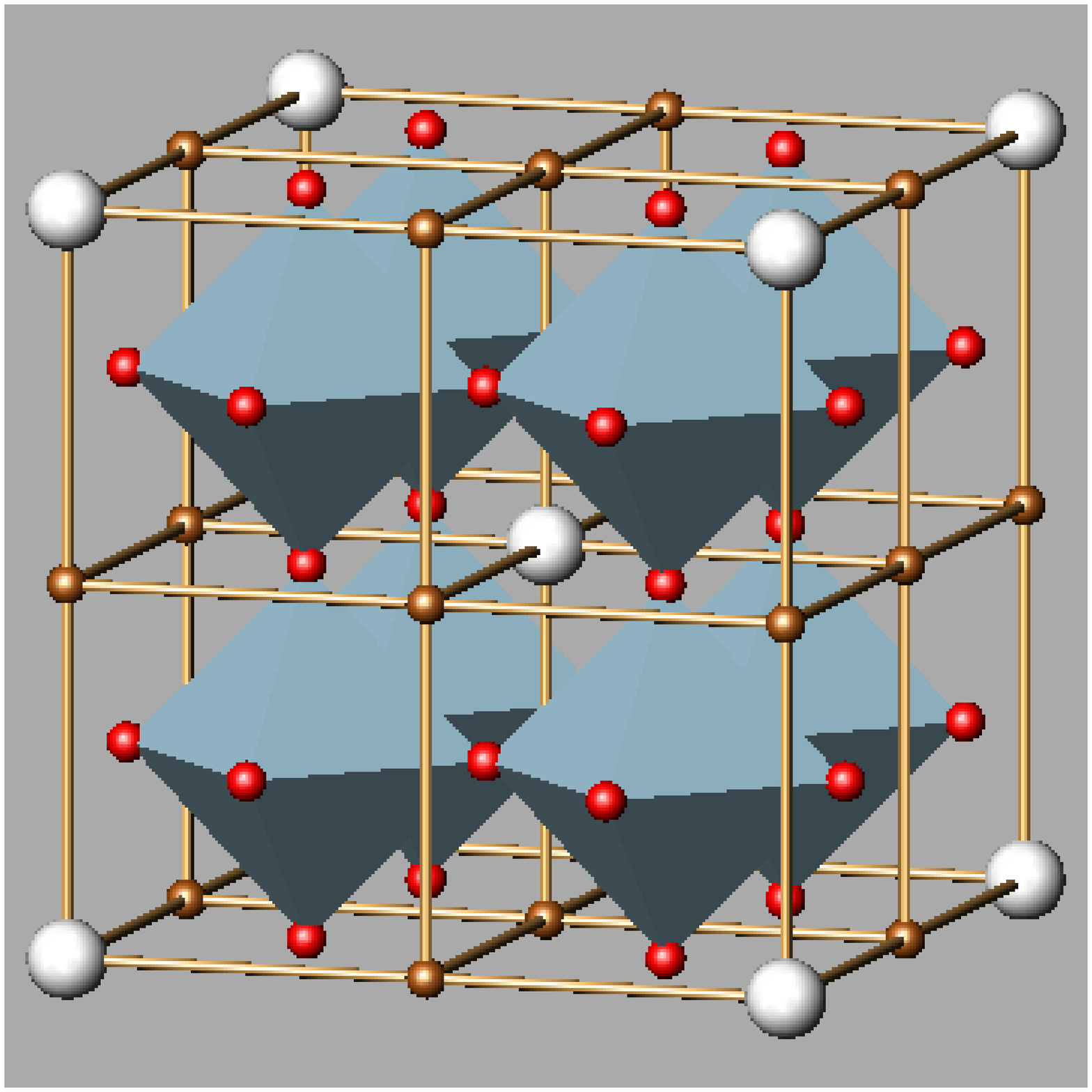}
\hspace{.3in}
\includegraphics[width=2.3in]{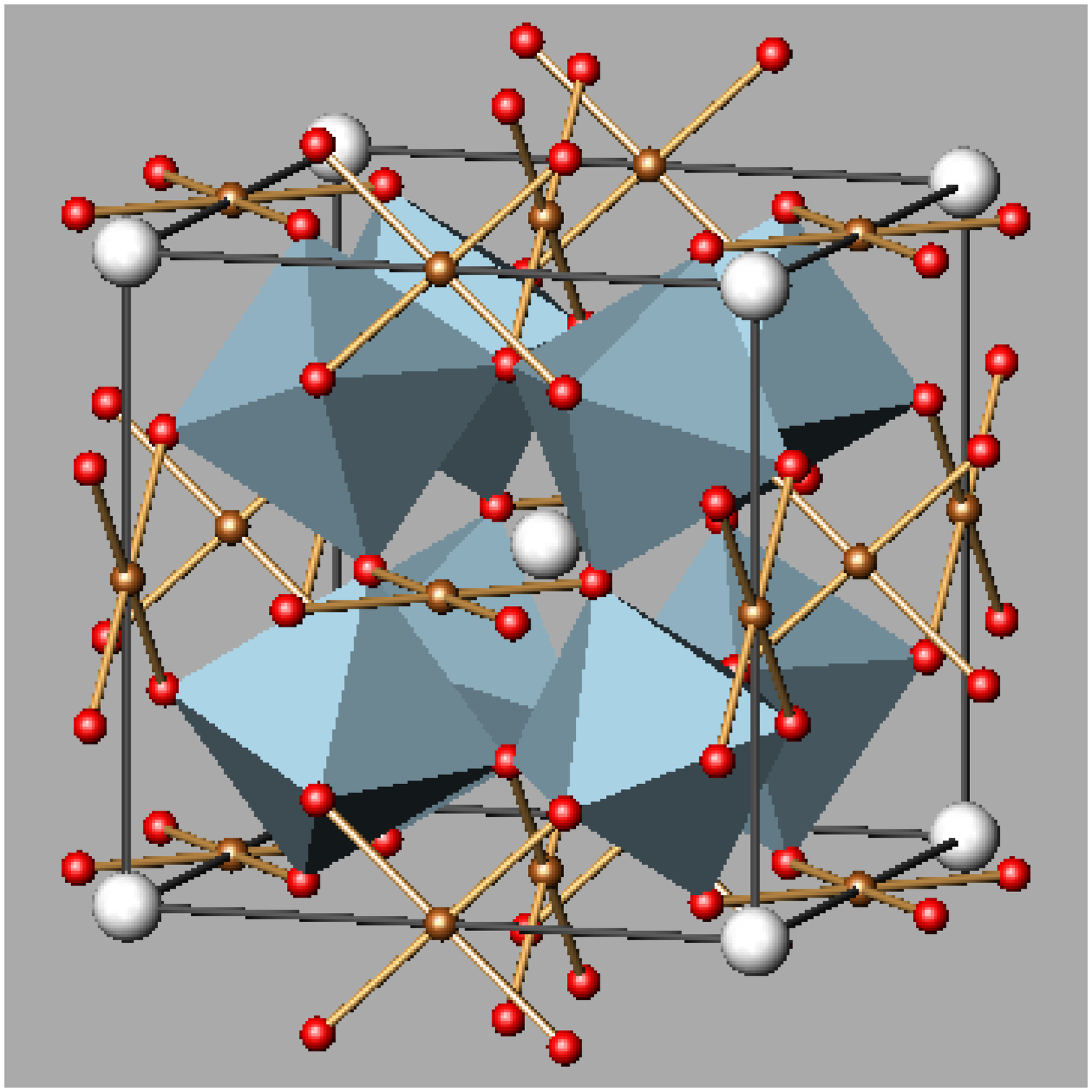}

\caption{ {\bf The quadrupled perovskite structure.} {\it  Top} The original perovskite structure is 
still 
obvious, despite the replacement of $\frac{3}{4}$ of the A-site ions.  {\it
Bottom}  
Tilted octahedra allow oxygens to form CuO$_4$ plaquettes with copper 
ions.}
\end{figure}

CaCu$_3$Mn$_4$O$_{12}$ (CCMO), a magnetic semiconductor,
combines the functional unit of the high temperature superconductors
-- the square CuO$_4$ plaquette -- with the functional unit of the colossal
magnetoresistance (CMR) manganites -- the MnO$_6$ octahedron -- in a
coherent (periodic) manner.  So far it is available only in polycrystalline
form, where it displays a substantial negative magnetoresistance (MR) beginning
from its magnetic ordering temperature of 370 K down to low temperature,
where its MR reaches -40\%.\cite{zeng}
Formally, two magnetic ions Mn$^{4+}$ ($d^{3},S=\frac{3}{2}$) and
Cu$^{2+}$ ($d^{9},S=\frac{1}{2}$) are anticipated, and magnetic ordering is
observed at 355 K.  The Cu and Mn magnetic moments are antialigned, 
resulting in a ferrimagnetic configuration.
The (polycrystalline) resistivity $\rho (T)$ is semiconducting in magnitude and
also in temperature derivative, but showed no visible anomaly 
upon ordering, \cite{zeng} so unlike in many manganites there is no insulator-metal
transition accompanying magnetic ordering.  The magnetization $M(T)$ shows a
steep, apparently first-order jump at T$_{C}$ to $\sim $80\% of its
saturation value. The temperature dependence of the activated resistivity
(over a limited temperature range) was fitted to obtain an 
energy gap (actually, an activation energy) of $\sim $0.12 eV.

CCTO has gathered recent interest due to its unusual dielectric properties.
At frequencies around 1 kHz, the extremely high dielectric constant
($\epsilon \approx$ 10$^4$) is nearly temperature independent between 25
and 300 C.\cite{exp,sinclair,science}   It is an antiferromagnetic (AFM)
insulator with an experimental gap of 1.5 eV as a lower limit. (see ref. 20 in \cite{he})
The AFM ordering is three-dimensional and has a N\'eel temperature of 25K.\cite{canted,koitzsch,kim}
The magnetic order of the Cu$^{2+}$ ions is N\'eel-like with each being
antialigned with its nearest neighbor on the (one-quarter depleted by Ca)
simple cubic lattice.  The Ti ions can have no net polarization by symmetry.
The formal valency of these ions is $d^0$, but our calculations show that some
amount of Ti $3d$ charge is present and is involved in magnetic coupling.

\section{Calculational Methods}

Calculations were carried out using the full-potential linearized 
augmented plane wave (FLAPW) method in the WIEN97 \cite{wien} implementation.  
The exchange-correlation potential of Perdew and Wang (92) \cite{xc} was used 
and over 2000 LAPW's were employed in the basis set.  In transition metal
oxides, there is often the question of whether corrections to the local density
approximation are necessary to obtain a reasonable description, with the 
answer being guided by experimental data.  While we suspect some corrections
will be necessary for both of these compounds, the results we have obtained
have the right magnetic character and are insulating, hence we expect they
are useful in beginning the interpretation of their observed behavior.
The experimental lattice 
parameter and ionic positions were used: 
for CCMO, $a$= 7.241\AA, $(y=0.3033,z=0.1822)$ for the O position;
for CCTO, $a$= 7.3843\AA, $(y=0.3033,z=0.1790)$ for the O position.
The crystallographic space group 
is Im3 but in order to reproduce the experimentally 
observed AFM order for CCTO, 
the unit cell had to be doubled.  
This results in a simple cubic cell; space group Pm$\overline{3}$.  The 
calculation was performed first using the fixed-spin moment procedure \cite{FSM} , with the moment held to zero.  Once a reasonable charge density was 
reached, the calculation was allowed to proceed to self-consistency with no 
restriction on overall moment.  The final moment indeed converged to zero and 
an anti-ferromagnetic ground state resulted.  For CCMO, the calculation was
begun with aligned Mn and Cu moments.  However, 
the Cu moment flipped during iteration, and only a
ferrimagnetic magnetic arrangement was obtained.\\
\vspace{-.3 in}
\section{CCMO}

\subsection{Electronic structure}

The calculated majority and minority band structures in the vicinity of the gap 
are shown in Fig. 2.  The
calculated energy gaps are 0.50 eV for spin up (parallel to the net
magnetization) and of 0.18 eV for spin
down, both direct. For the minority carriers the gap occurs at H
between pure Cu $d_{xy}$ states below the gap to Mn $t_{2g}$ character
above. The small thermal gap, 0.09 eV, is {\it indirect} between
the spin down valence band maximum at H and the spin up conduction band
minimum at $\Gamma$. It is common for the band gap in density
functional calculations to be
smaller than the true gap, but the band character and shape on either side
of the gap typically are given reasonably. In this case the calculated gap
is quite similar to the quoted experimental value.

\begin{figure}
\includegraphics[height=2.5in]{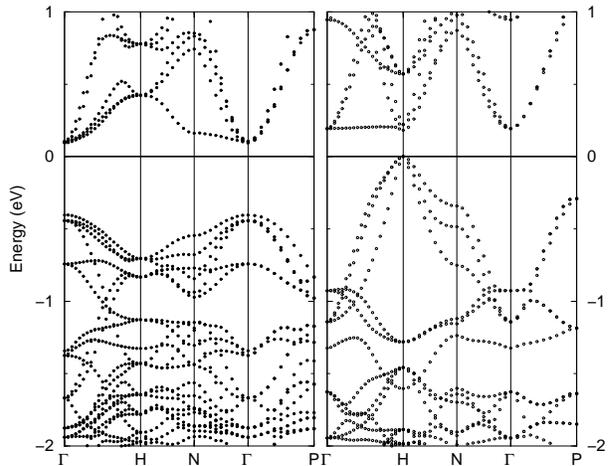}
\caption{{\bf Bandstructure of CCMO in the region of the fundamental gap.} The indirect thermal gap is between the spin-down states at the H point (right panel) to the spin-up states at $\Gamma$ (left panel)}  

\end{figure}

As in many related cuprates, the hole in the Cu $3d$ shell (the magnetic 
orbital) involves the $dp\sigma$ antibonding combination of the Cu d$_{xy}$ and
the four neighboring O $p\sigma$ orbitals, a Wannier function 
which we denote as ${\cal D}_{xy}$.
The other Cu $d$ orbitals are concentrated in a small region between 1 and
2.5 eV below the gap.
If the CuO$_4$ unit were actually a square, ${\cal D}_{xy}$ would not couple
with other Cu $d$ -- O $p$ combinations, and this is a good approximation.
From the structural figures, it is clear that
${\cal D}_{xy}$ functions on
both near neighbor and second neighbor Cu are orthogonal by symmetry.
As a result, the dispersion involves further neighbors as well as Mn-centered
Wannier functions, and becomes difficult to model with a tight-binding approach, (see the sections on CCTO, below).\\

\subsection{Magnetic ordering and character}

The calculated moments inside the muffin tin spheres are
2.42 $\mu_B$ and -0.45 $\mu_B$ for Mn and Cu respectively.
The idealized values ($S=\frac{3}{2}$, 3 $\mu_B$ for Mn and
$S=\frac{1}{2}$, 1 $\mu_B$ for Cu) are reduced due to 
hybridization with the O 2$p$ states, as observed in other transition
metal oxides.
The net magnetic moment of 9 $\mu_B$ per formula unit is however
what would be obtained
from the formal moments aligned ferr{\it i}magnetically.
The magnetic moment on the Cu ion is a result of the exchange splitting and resulting single occupation of
the ${\cal D}_{xy}$ orbital.  Magnetism in the Mn ion comes as expected from the
splitting of the t$_{2g}$ orbitals and filling only of the spin up states.

The magnetism of CaCu$_3$Mn$_4$O$_{12}$ presents several interesting
questions. First of all, the exchange splitting of the Cu ${\cal D}$ orbital is
$\Delta^{Cu}_{ex} \sim$ 1.2 eV.  
Although such exchange splittings are less
apparent in ferrimagnets than in simple ferromagnets, this Cu $d$ orbital
is separate from the other Cu $d$ orbitals, making it
easy to identify the splitting from the density of states.  Given its moment of
$m_{Cu}$ = 0.45 $\mu_B$ (inside its muffin tin sphere), there results a ratio 
$\Delta^{Cu}_{xy}$/$m_{Cu}$ = 2.7 eV/$\mu_B$,
a very large value considering this ratio (the Stoner $I^{Cu}_{xy}$
in the relation $\Delta_{ex} = I m$)
is usually no more than 1 eV/$\mu_B$ in transition metal magnets. The exchange
splitting of the other Cu $d$ orbitals is smaller, of the order of 0.4 eV.
This difference reflects a strongly anisotropic exchange potential on the Cu
ion, which bears further study.

The magnetic coupling in CaCu$_3$Mn$_4$O$_{12}$~ is potentially quite
complex to unravel due to the presence of two types of magnetic 
ions and the large cell
with low symmetry sites. The various bond angles involved in the exchange
processes in CaCu$_3$Mn$_4$O$_{12}$~ have been discussed in our earlier
paper,\cite{CCMOprb} and there may be alternative 
insights to be gained first -- these
are addressed below in the discussion of CCTO.
Furthermore, Mn-Mn coupling is involved as well as the Mn-Cu coupling
discussed in the previous section. Since Mn spins lie on a simple cubic
lattice connected by a single O$^{2-}$ ion, the Goodenough-Kanamori-Anderson
(GKA) rules can be applied to understand $J^{Mn-Mn}$ of ferromagnetic
sign. Further Mn-Mn
couplings should be small and not affect qualitative behavior. Whereas 180$%
^{\circ}$ Mn$^{4+}$-O-Mn$^{4+}$ coupling is antiferromagnetic ({\it viz.}
CaMnO$_3$), when this angle is reduced to 142$^{\circ}$ a ferromagnetic sign
may result from the GKA rules. Parallel alignment of the Mn spins is observed, and is the only
situation we have encountered in our calculations.
Finally, there is the question of the origin of the AFM
Mn-Cu coupling. In this structure all
oxygen ions are equivalent and each one is coordinated 
with two Mn ions and one Cu
ion.   As discussed elsewhere,\cite{CCMOprb} if only nearest neighbor
Mn-Cu coupling is appreciable, it is AFM in sign and $\approx$ 20 meV.  Cu-Cu coupling in this structure is discussed in the next section.\\

\subsection{Transport behavior.}

Several unusual aspects of transport and of 
thermally-induced carriers in CCMO have been discussed previously.  The
main experimental interest so far has been in the magnetoresistance, 
becoming appreciable as it does above room temperature and continuing
(indeed, increasing) to low temperature.\cite{zeng}  Due to the 100\% spin polarization
of thermally induced or doped-in carriers in a spin-asymmetric semiconductor
such as we predict CCMO to be, intergrain transport is very strongly dependent
on the direction of magnetization of neighboring grains.  The effect is
directly analogous to arguments applied to granular half-metallic ferromagnetic
materials.  The consequences have been discussed by several authors; see in
particular the review by Ziese.\cite{ziese}  A magnetoresistance of 30-40\%,
as seen in CCMO at low temperature, is readily accounted for by the alignment
by the field of grains with fully polarized carriers.\\

\section{CCTO}

Our calculation correctly reproduces 
the observed AFM insulating character of CCTO.  The gap in the bandstructure, shown in Fig.3 is only 0.2 eV, 
considerably smaller than the experimental value, but such 
underestimation of the gap is common in LDA calculations and arises
from underestimation of correlation effects. Our density of states is similar to that presented by He et al. \cite{he}
The band structure reflects the six bands, associated with six Cu ions,
that are responsible 
for the magnetism of the compound.  These states are the same ${\cal D}_{xy}$
Wannier orbitals introduced in the discussion of CCMO above.  Since these six
bands are separated from other bands,
they are good candidates for providing a minimilist representation of the 
origin of magnetic coupling. The bands are fit to a tight-binding model 
and the hopping amplitudes are used to obtain the spin-spin coupling $J = 
\frac{4t^2}{U}$ via the superexchange mechanism.  This prescription has 
worked well in several structurally complicated magnetic insulators such 
as CuV$_4$O$_9$, \cite{hellberg}  Li$_2$CuO$_2$, \cite{weht} and 
LiVOSiO$_4$. \cite{rosner}   Sometimes, however, a multi-orbital model 
may be required. \cite{feldkemper} An analysis 
of the character of these half-filled bands shows that they contain some 
Ti $3d$ character
in addition to the expected Cu $d$, O $p$ antibonding character, and therefore 
that the Ti (nominally $d^0$ ion) is involved in the magnetic coupling. 

\begin{figure}[tbs]
\includegraphics[height=2.5in]{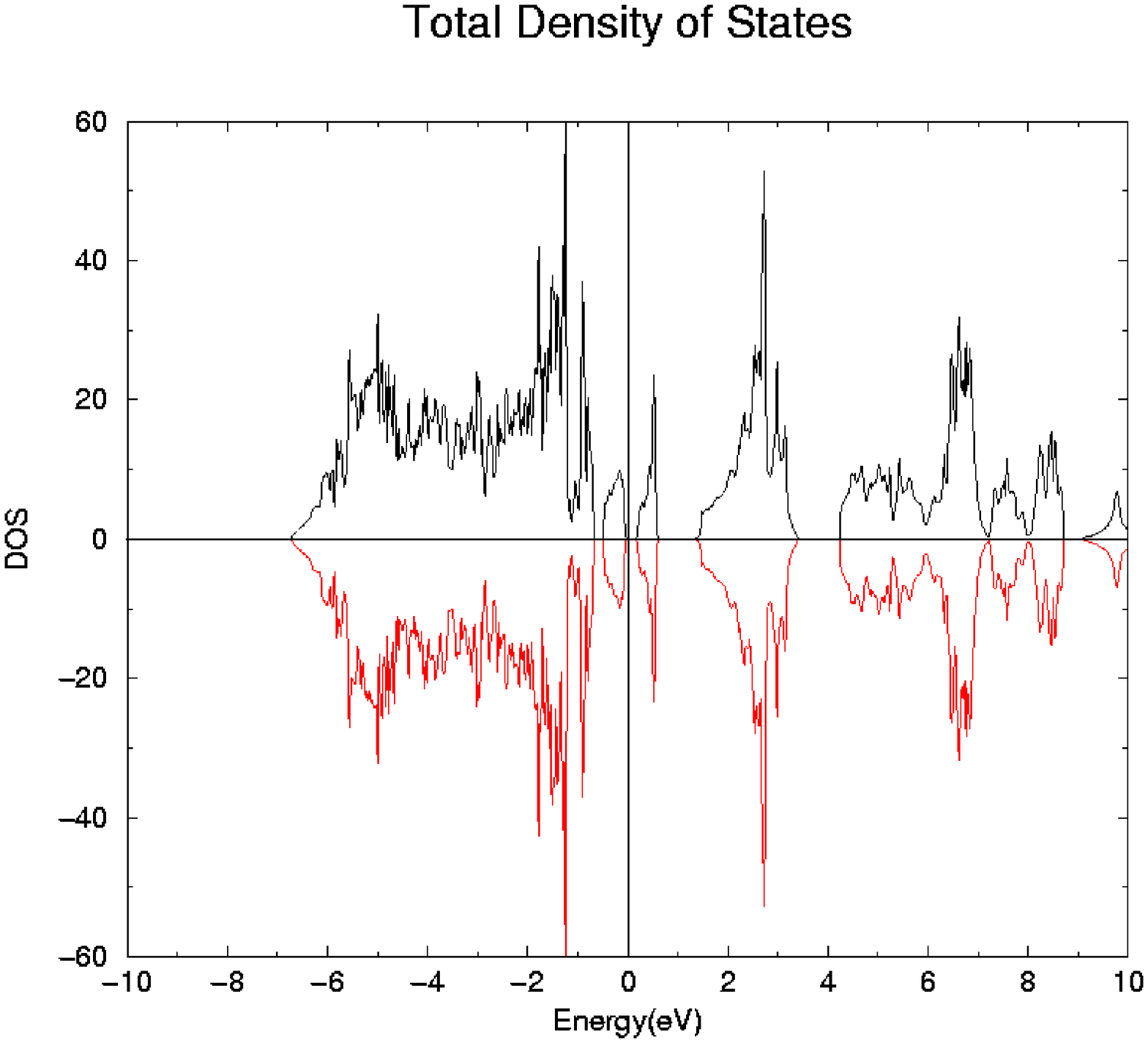}
\hspace{.5in}
\includegraphics[height=3.0in]{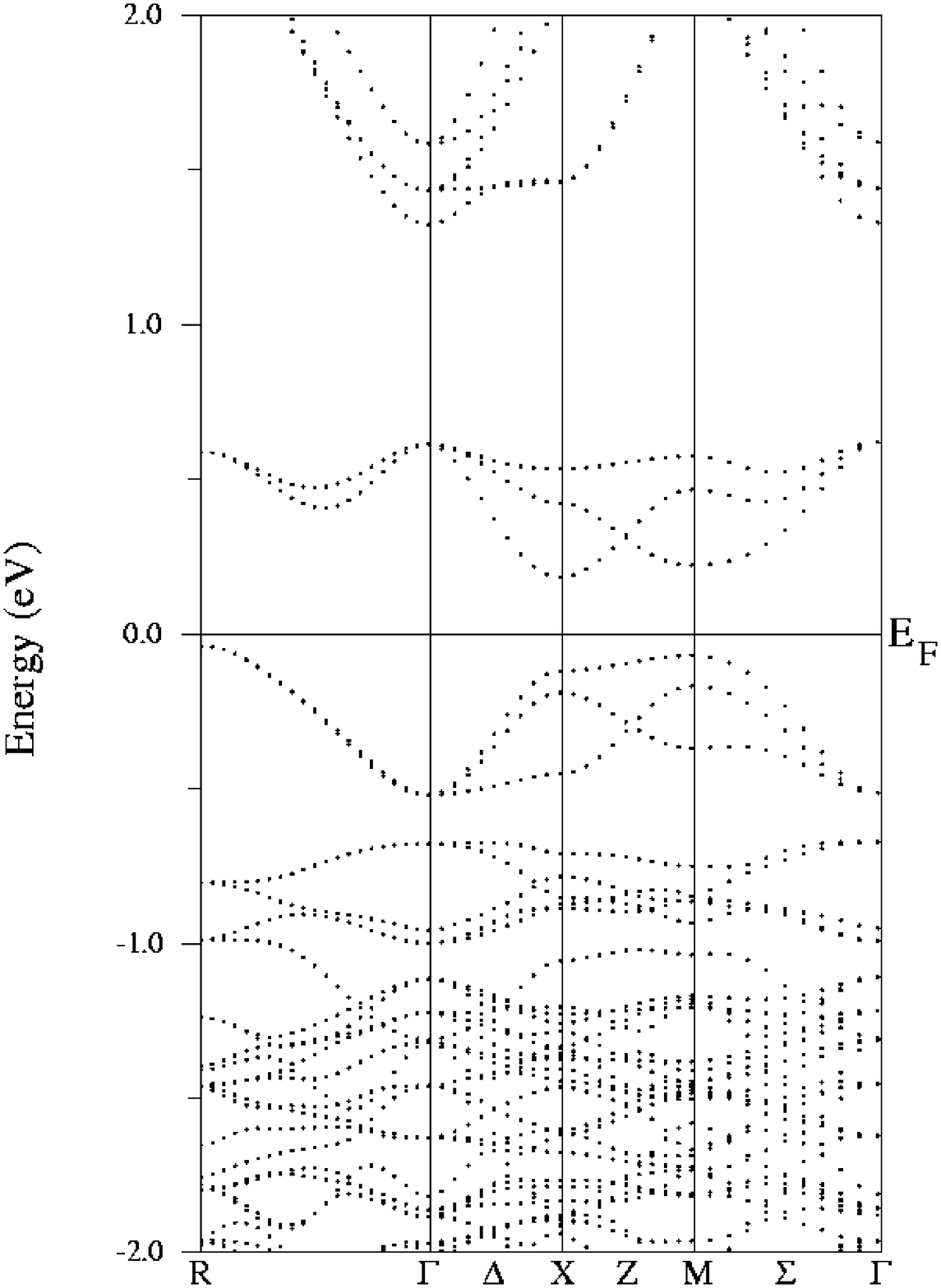}
\caption{ {\bf CCTO Density of States and Bandstructure.}{\it Top} The small gap indicates an insulating 
ground state and the identical spin-up and spin-dn densities are consistent 
with AFM ordering. {\it Bottom} The six magnetic bands are separated from all others.  Three are filled 
and three are unfilled. }
\end{figure}

\subsection{Tight-Binding Model.}

The most basic tight-binding model includes only the $\mathcal{D}_{xy}$ Wannier 
functions centered on each of the six Cu sites in the AFM cell.
The Wannier functions on different sites are 
orthogonal, and they have the symmetry of the Cu $d_{xy}$ orbital, which provides
essential selection rules.
The contributions of many sites are present due to hybridization, and for CCTO 
contains Ti $3d$ character, as mentioned above.  The hopping parameters may be 
fit from the bandstructure, in fact, for CCTO it has been possible to use the
paramagnetic band structure.  

Typically only 1st and 2nd nearest neighbor hopping parameters are 
required to describe the band dispersion.  However, for CCTO the 1st and 2nd 
Cu-Cu hopping amplitudes (hence superexchange coupling, see below) vanish by symmetry\cite{SK}.  
The 3rd neighbor interactions are non-zero and reproduce 
the general band dispersion, but important degeneracies are not broken. 
This can be understood by noticing that 
CCTO is made up of three separate BCC sublattices, each containing co-planar 
CuO$_4$ squares with their associated Wannier functions.  The 3rd neighbor 
interactions couple only Cu sites within 
a given sublattice, hence the bands are degenerate at this level of description.  
Hopping between Cu sites belonging to two 
non-co-planar CuO$_4$ squares is necessary to split the bands.  
The 4th nearest neighbor interaction is 
non-zero but, remarkably, again couples only Cu sites within a single sublattice.   
Only at 5th neighbor is there non-zero hopping between  
different sublattices.  This hopping involves Cu sites which are 
separated by more than a lattice constant (which itself is twice a conventional
perovskite lattice constant!) but it is essential in order to obtain hopping (and therefore superexchange coupling) between the three sublattices.  
To reproduce the band structure accurately, it is necessary to include 
7th neighbors.  Although this 7th neighbor interaction again couples 
Cu sites only within each sublattice, it clearly contributes non-negligibly to the 
observed dispersion.  Including all non-zero interactions up to 7th
neighbor yields a very accurate bandstructure. (Fig.4)
\begin{figure}
\begin{center}
\includegraphics[width=2.2in,angle=270]{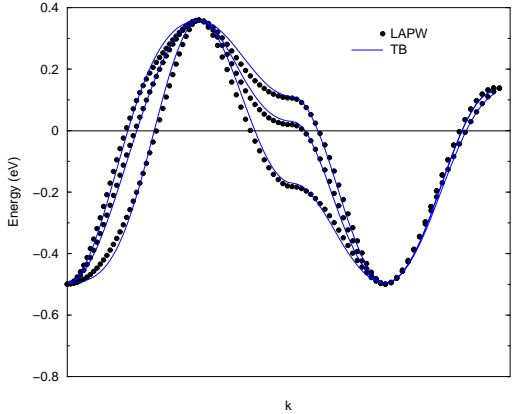}
\end{center}
\caption{ {\bf LAPW vs. Tight-Binding Bandstructure.}  The tight-binding fit includes up to 7th neighbor parameters.  This produces a bandstructure remarkably similar to the all-electron LAPW calculation. }
\end{figure}
The hopping parameters used to make this fit are as follows:

\begin{center}
\large
\begin{tabular}{|ll|r|} 
\multicolumn{2}{r}{} &
 \multicolumn{1}{l}{} \\ \hline
t$_{3rd}$ & & -53.0 meV   \\ \hline
t$_{4th}$ & $\pi$(Ca) & -51.0 meV  \\ \hline
t$_{4th'}$ & $\pi$(no Ca) & 16.3 meV \\ \hline
t$_{4th''}$ & $\delta$ &  20.1 meV \\ \hline
t$_{5th}$ & & 5.6 meV \\ \hline
t$_{6th}$ & & -.7 meV \\ \hline
t$_{7th}$ & $\sigma$ & -1.4 meV  \\ \hline
t$_{7th'}$ & $\delta$ (Ca) &  2.7 meV \\ \hline
t$_{7th''}$ & $\delta$ (no Ca) &  -9.5 meV \\ \hline
\end{tabular}
\end{center}

For both 4th and 7th neighbors, there are several different parameters.  This is for one of two reasons. In some cases, the  Wannier orbitals 
involved in these hopping terms are oriented differently with respect to one 
another along different directions (i.e. towards different neighbors of the 
same distance)  which produces a different magnitude of hopping.  In other 
cases, the chemical environment is actually different, such that the path from 
one orbital to another along one direction requires going through a Ca while 
along a second direction it does not.  The orientation differences are indicated 
using the common chemical bonding labels and the chemical differences are noted 
by stating whether or not Ca ions are involved.  The little information for exchange coupling in CCTO to date is from magnetic Raman spectroscopy.  Koitzsch et al. \cite{koitzsch} fit magnetic excitations to 1st, 2nd, and 3rd neighbor exchange couplings, which allows no reasonable comparison with our values.  Linear spin wave calculations are in progress that will allow closer comparison with experiment.\\
\subsection{Magnetic Order}

Since the Cu ion has spin half, there is no first order single ion anisotropy and the spin behavior should be Heisenberg-like.  Using the tight-binding hopping parameters, the Heisenberg model based on
superexchange coupling may be formulated, the notation being standard:
\begin{equation}
H = \sum_{<i,j>} J_{ij}\vec{S_i} \cdot \vec{S_j}\ \hspace{.3in} J_{ij} = \frac{4t_{ij}^2}{U}
\end{equation} 

The on-site repulsion U is taken to be 4.0 eV, as similar values have been used for such ``single band''
representations of cuprates.  The magnetic order that results from 
applying this model has no dependency on the value of U, except that it 
be large enough to justify the use of the Heisenberg model (t$^2$ $\ll$ U).
The value of U ({\it i.e.} J) will however affect the ordering temperature.  
Regardless of the sign of t$_{ij}$, J$_{ij}$ from this superexchange treatment
will be positive (favoring opposite alignment of the two spins).

Using this observation, it is clear that J$_{3rd}$ will anti-align a given spin 
with its third nearest neighbor, as is observed.   However, as noted above, this coupling is only between spins on the same sublattice so although it results in sublattice ordering, there is no alignment of sublattices with respect to one another.  The next coupling, J$_{4th}$ wants to anti-align the spin with its 4th nearest neighbor.  However, 4th nearest neighbors share a third neighbor with which they are already anti-aligned.  The system is therefore frustrated.  The largest J$_{4th}$ is slightly smaller than J$_{3rd}$ and there are only two such neighbors as compared to the eight of J$_{3rd}$. Thus at this level, all sublattices are antialigned internally, though with some frustration.  Intersublattice coupling occurs because J$_{5th}$ antialigns a given spin with its 5th neighbor, in effect antialigning nearest neighbors, thereby correctly reproducing the three-dimensional anti-ferromagnetic order of CCTO.  In this picture, then, full three-dimensional AFM order depends on interactions as far away as 5th Cu neighbors, which are actually  separated by many more than five ions. 
\vspace{-.35in}

\section{Conclusions}
Ferrimagnetic order has been calculated for CCMO with total moment of 9$\mu_B$ per formula unit, consistent with 
experiment.  The predicted spin-asymmetric gap leads to 100\% spin 
polarized carriers that may be essential in accounting for its large and 
weakly temperature dependent magnetoresistance.
The superexchange picture of magnetism has been applied to CCTO, 
and suggests that extremely long-range 
interactions are important and perhaps even responsible for the AFM order.   Within this model, there is an energy advantage (for classical spins) of  $\sim$ 23 meV/Cu favoring AFM over FM alignment. Though mechanisms besides superexchange may be important to fully understand the magnetism of this compound, this simple picture is able to account for the experimentally observed order. \\
\vspace{-.32 in}

\section{Acknowledgments}{

We are grateful to C.C. Homes, A.P. Ramirez, S.M. Shapiro, and M. Subramanian for discussions and communication of unpublished work.  We benefitted greatly from input on the tight-binding procedure for CCTO from Wei Ku.  This work was supported by NSF grant DMR-0114818.\\


\end{document}